# Reduced-Dimension Surrogate Modeling to Characterize the Damage Tolerance of Composite/Metal Structures

**Corey Arndt, Cody Crusenberry, Bozhi Heng, Rochelle Butler and Stephanie TerMaath \***

Department of Mechanical, Aerospace, and Biomedical Engineering, The University of Tennessee, Knoxville, TN 37996, USA
\* Correspondence: stermaat@utk.edu

**Abstract:** Complex engineering models are typically computationally demanding and defined by a high-dimensional parameter space challenging the comprehensive exploration of parameter effects and design optimization. To overcome this curse of dimensionality and to minimize computational resource requirements, this research demonstrates a user-friendly approach to formulating a reduced-dimension surrogate model that represents a high-dimensional, high-fidelity source model. This approach was developed specifically for a non-expert using commercially available tools. In this approach, the complex physical behavior of the high-fidelity source model is separated into individual, interacting physical behaviors. A separate reduced-dimension surrogate model is created for each behavior and then all are summed to formulate the reduced-dimension surrogate model representing the source model. In addition to a substantial reduction in computational resources and comparable accuracy, this method also provides a characterization of each individual behavior providing additional insight into the source model behavior. The approach encompasses experimental testing, finite element analysis, surrogate modeling, and sensitivity analysis and is demonstrated by formulating a reduced-dimension surrogate model for the damage tolerance of an aluminum plate reinforced with a co-cured bonded E-glass/epoxy composite laminate under four-point bending. It is concluded that this problem is difficult to characterize and breaking the problem into interacting mechanisms leads to improved information on influential parameters and efficient reduced-dimension surrogate modeling. The disbond damage at the interface between the resin and metal proved the most difficult mechanism for reduced-dimension surrogate modeling as it is only engaged in a small subspace of the full parameter space. A binary function was successful in engaging this damage mechanism when applicable based on the values of the most influential parameters.

**Keywords:** finite element analysis; reduced-dimension surrogate modeling; sensitivity analysis; layered structure; damage tolerance; surrogate modeling

## 1. Introduction

Computational simulation using validated, high-fidelity, physics-based models is an effective method to evaluate complex engineering designs that encompass multiple, interacting physical behaviors. Such complex models are typically computationally demanding and defined by a high-dimensional parameter space challenging the comprehensive exploration of parameter effects and design optimization. As such, the computational time and memory requirements to generate an adequately sampled dataset of potential parameter combinations for design characterization counteract the benefits of computational exploration, even when utilizing a high-performance computing (HPC) system.

One means of reducing this computational effort is to formulate a surrogate model. A surrogate model is a fast-running approximate tool that represents the trend between the input variables and the response quantities of interest within a specified level of accuracy. As categorized by Eldred et al. [1], the three classes of surrogate models are reduced-order, hierarchical, and data fit models. Surrogate modeling has been demonstrated in many engineering fields including fluid dynamics, composite layered structures, aerospace, and others [2-5]. For example, Peherstorfer and Willcox [2] presented a surrogate method to simplify the parameter space controlled by partial differential equations (PDEs) and dynamic sensor data. They successfully reduced the complexity of discretizing the PDEs with the use of models that adapted to the real-time sensor data. This was achieved with proper orthogonal decomposition (POD) which represents a given set of data in a least-squares optimal sense. Vendl and Faßbender [5] also utilized POD to generate least-squares surrogate models of digital flight tests to shorten the computationally exhaustive numerical simulations usually required to collect aerodynamic representations of flight data. Kiani et al. [6-9] proposed integro- and differential-nonlocal-based models to evaluate the surface energy on nanostructures to identify internal and interfacial defects. Kiani et al. utilized these models to study local defect effects on various aspects of vibrations and frequencies including longitudinal, torsional, and twisting. Further examples of reducing the complexity of the damage mechanisms to analytical representations have been presented by Mousavi et al. and Jassim et al. [10,11]. These studies investigated the identification of damage mechanisms and defects through the use of reduced complexity analytical equations such as Variational Mode Decomposition, Empirical Mode Decomposition, Frequency Reduction Index, and Modal Assurance Criteria. Wu et al. [3] produced surrogate models of the elastic response of representative volume elements (RVEs) for the multi-scale components of a woven composite material. They did not include damage mechanisms and the model was limited to a single 2D woven fabric. While effective, these methods are all advanced approaches requiring substantial expertise and/or user effort to implement.

Of the three types of surrogate modeling, the data fit method is the most user friendly for a non-expert. Surrogate modeling by fitting data uses a limited number of source model (the original high-fidelity physics-based model) evaluations to develop a non-physics-based mathematical model that serves as an alternate formulation to capture the phenomena of interest. As cautioned by Frangos et al. [12], this approach is a black box, therefore it is essential to understand the consequences of methodology choices when formulating the surrogate model. Numerous mathematical formulations have been employed for surrogate modeling ranging from simple linear/nonlinear regression and polynomial response surfaces to advanced learning algorithms such as artificial neural networks [13]. Reduced-dimension surrogate modeling allows for further simplification by formulating a surrogate model using only the most influential input parameters as variables. In this paper, this type of mathematical representation using surrogate models formulated through data fitting is called a reduced-dimension surrogate model (RDSM).

The goal of this research is to demonstrate an approach to RDSM formulation using commercially available tools requiring minimal expertise. In this approach, the complex physical behavior of the high-fidelity source model is separated into individual, interacting behaviors. A separate RDSM is created for each behavior and then all are summed to formulate the RDSM of the source model. In addition to substantial reduction of computational effort, this method also provides a characterization of each individual behavior providing insight into the source model. The approach encompasses experimental testing, finite element analysis, surrogate modeling, and sensitivity analysis and is demonstrated by formulating a RDSM for the damage tolerance of a layered structure (an aluminum plate reinforced with a co-cured bonded E-glass/epoxy composite laminate) under four-point bending. This problem was chosen because multiple damage mechanisms contribute to the structural failure, and these damage mechanisms are governed by a multitude of parameters resulting in a nonlinear, high-dimensional, interacting parameter space, thus challenging structural characterization and efficient damage prediction.

A separate RDSM is created for the energy absorbed by each damage mechanism individually (mechanism RDSM) and these formulations are then summed to create a source RDSM to predict the total damage energy absorbed by the layered structure. In total, five mechanism RDSMs are produced with the use of surrogate models on the reduced parameter space for each damage mechanism as determined by sensitivity analysis. Energy was chosen as the output because it is readily extracted from finite element (FE) results and its value can be converted into a quantifiable damage measure. In addition to ease of implementation in exploring the total damage tolerance of the layered structure, two other outcomes are achieved with this approach. A primary advantage is that each damage mechanism is characterized individually, providing an understanding of its contribution to damage and the parameter ranges affecting the magnitude of damage. These parameter ranges define the subspace in which the mechanism contributes to the total damage energy. Secondly, the most influential parameters on damage tolerance are identified, enabling focused optimization and analysis limited to these parameters. This information also enables tailored material selection to produce particular modes of damage initiation and informed inspection to identify non-visible internal damage. The summed RDSM approach can be generically applied to varying loading conditions, geometric configurations, and boundary conditions.

## 2. Materials and Methods

The source model for the layered structure is a high-fidelity 3D finite element (FE) model, previously developed by Heng et al. and Arndt et al. [14-17]. The structural configuration is an E-glass/epoxy composite overlay co-cured to an aluminum 5456-H116 substrate (Figure 1). This source model captures five interacting damage mechanisms including matrix cracking, fiber fracture, delamination within the composite, disbond at the composite/metal interface, and yielding in the metal. Experimental test results were used to validate the computational model. In addition to force and displacement measurements, Digital Imaging Correlation (DIC) provided high-resolution data on damage propagation and the strain field. The damage tolerance of the composite/metal layered structure was investigated under four-point bending, a loading condition that activates all five damage mechanisms. To demonstrate the summed approach, a mechanism RDSM is formulated separately for each damage type and these mechanism RDSMs are summed to predict the total damage. This summed approach is compared to a traditional, direct approach that formulates a RDSM directly from the total damage energy obtained from the source model (Figure 2). It will be shown that the summed RDSM better captures damage behavior throughout the entire parameter space, is faster to formulate, and provides more information. As described below, the RDSM formulation approach integrates experimental testing, computational modeling, surrogate modeling, and sensitivity analysis.

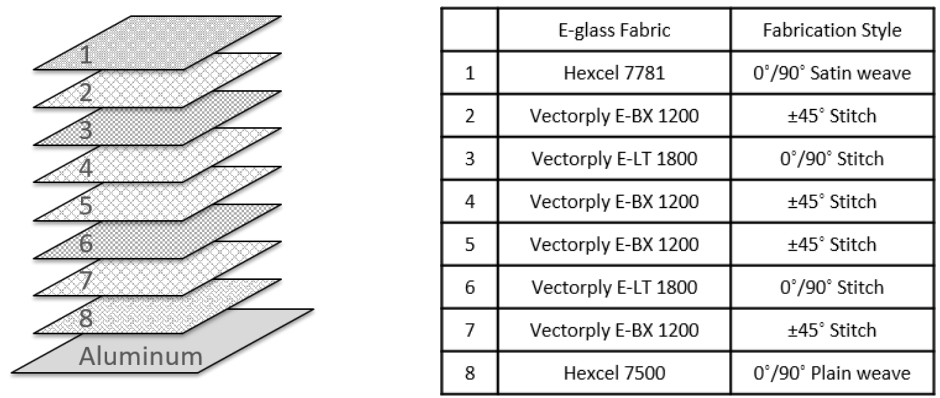

| | E-glass Fabric | Fabrication Style |
|---|---|---|
| 1 | Hexcel 7781 | 0°/90° Satin weave |
| 2 | Vectorply E-BX 1200 | ±45° Stitch |
| 3 | Vectorply E-LT 1800 | 0°/90° Stitch |
| 4 | Vectorply E-BX 1200 | ±45° Stitch |
| 5 | Vectorply E-BX 1200 | ±45° Stitch |
| 6 | Vectorply E-LT 1800 | 0°/90° Stitch |
| 7 | Vectorply E-BX 1200 | ±45° Stitch |
| 8 | Hexcel 7500 | 0°/90° Plain weave |

**Figure 1.** Composite/metal layered structure configuration.

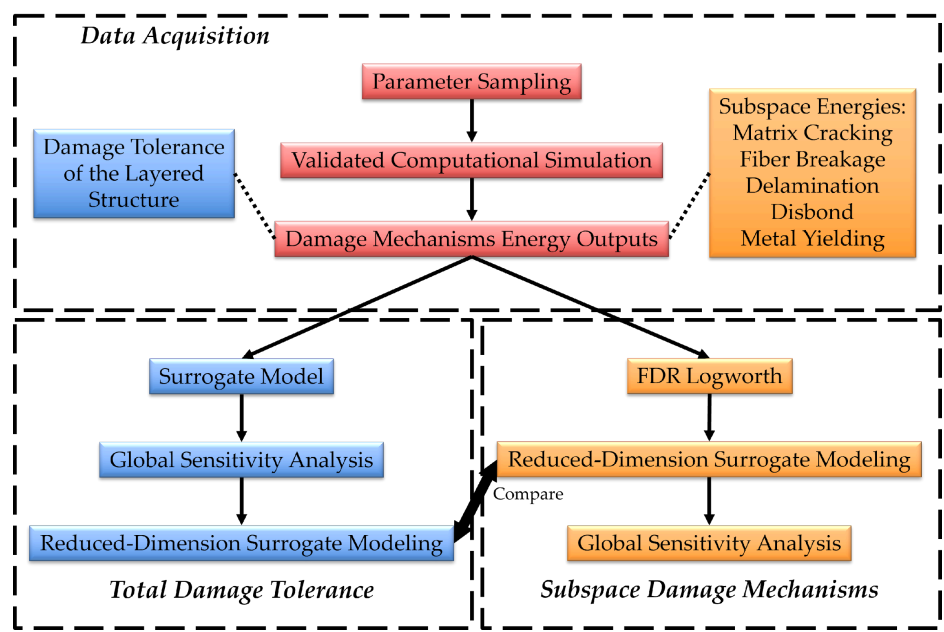

**Figure 2.** Flowchart comparing the direct and summed RDSM approaches.

## 2.1. Composite Overlay Background

Fiber-reinforced composite overlays are installed to reinforce or repair metallic structures in engineering fields spanning aerospace [18-23], marine [24,25], automotive [26], and infrastructure [27,28]. Composite overlays offer many advantages over traditional repair methods (e.g., mechanical fastening of a patch or structure replacement), including uniform stress transfer, easy installation, customized stiffness, high specific strength, adaptability to complex substrates, and excellent corrosion resistance [20]. However, the resulting layered structure presents complex, progressive damage that challenges optimized design, prediction of structural reliability, and informed inspection/non-destructive evaluation. This progressive damage occurs externally and internally (non-visible) under design loads and overloads, and given the thickness of the aluminum substrate in many applications, bending effects must be considered [21,29-32]. Damage progresses as matrix cracking, fiber fracture, delamination within the composite, disbond at the composite/metal interface, and plastic deformation.

Matrix cracking is observed when the resin that transfers load to the fibers and supports the fibers in compression fails. Fibers can fracture in tension or compression. Delamination is the result of lamina plies peeling apart when the resin interface fails.

Disbond is the same damage type as delamination and occurs at the composite/metal interface. Yielding in the metal presents itself as plastic deformation of the substrate. The shortened notation provided in Table 1 is used to describe these five damage mechanisms. A 2-letter system indicates the type of energy dissipated and the damage location. The energy types are plastic (P), damage (D), and total (T). There are four locations: lamina (L), cohesive layers between lamina plies (C), the interface between the composite and metal (I), metal substrate (M), and the structure in its entirety (S). These varying damage mechanisms are controlled by a multitude of parameters and damage mechanisms resulting in a complex, high-dimensional, interacting parameter space that governs the damage tolerance of this layered structure.

**Table 1.** Notations for damage energy outputs.

| Damage Mechanism | Subspace | Notation |
|---|---|---|
| Matrix Cracking | Shear Plasticity | *PL* |
| Fiber Fracture | Intralaminar Fracture | *DL* |
| Delamination between the Laminae | Delamination | *DC* |
| Disbond at the Composite/Metal Interface | Disbond | *DI* |
| Plastic Deformation in the Metal | Plastic Deformation | *PM* |
| Damage Tolerance of the Layered Structure | Total Energy Absorbed | *TS* |

There is a considerable amount of prior research on the damage analysis of composite overlay repair focusing on specific damage mechanism topics such as the fracture behavior of bonded metallic substrates [33-37] and adhesive disbond [38-42]. It has also been demonstrated that the damage within the composite overlay can substantially reduce the efficiency of the reinforcement or repair [43,44]. Moreover, Jones [43] concluded that multiple failure modes, including cracking in the adhesive or at the adhesive–metal interface, fiber fracture, and delamination, should be evaluated when performing damage tolerance assessments of layered structures. The progressive failure resultant from these damage mechanisms was subsequently investigated (e.g., [45-51]). For example, an aerospace application of composite/metal layered structures occurs at stiffener to skin bonding, i.e., Pi and T joints. Action et al. [50] investigated the fail-safe crack arrestment in a T -joint bonded to aircraft skin. They studied crack arrestment for a multi-loaded bonded T joint looking at strain energy to predict the delamination and crack propagation. Finlay et al. [51] applied a cohesive zone model to investigate the progressive failure of composite Pi joints for pull-off and side-bend loading. Their research expanded upon 3D joint models developed by Novak and Selvarathinam [52] who utilized XFEM to study both inter-laminar delamination and intra-laminar damage. While these studies investigated the progressive damage of composite/metal layered structures, a method is needed to characterize the high-dimensional, complex, interacting parameter space encompassing the varying damage mechanisms. This knowledge is essential to prevent damage initiation, plan inspections particularly for internal damage, design for customized failure to provide ample time for detection and repair, and to optimize the design for weight, safety, and performance based on material selection.

### 2.2. Layered Structure Fabrication

The composite/metal layered structure was fabricated by installing an E-glass/epoxy composite overlay onto a 0.25-inch-thick aluminum 5456-H116 plate using a hand lay-up procedure and vacuum infusion. The bonded side of the aluminum plate was prepared by polishing and cleaning, followed by the application of AC-130. A quasi-isotropic laminate was produced with ±45° (Vectorply E-BX 1200) and 0°/90° (Vectorply E-LT 1800) stitched fabrics as the main fabric plies. A 0°/90° plain weave fabric (Hexcel 7500) was selected for the resin-rich side bonded to the aluminum substrate. A 0°/90° fine harness stain weave fabric (Hexcel 7781) was selected for the top side to obtain a quality surface. The stacking sequence is provided in Figure 1 for the 0.16-inch-thick composite overlay. An epoxy resin (M1002) was mixed with a curing agent (M2046 hardener) for the composite matrix and adhesive layer. The resulting wet composite overlay was covered by a P3 perforated film to control bleed and then covered with a vacuum bag. The vacuum level was set to 20 inHg for 3 h and then the layered structure was cured in an oven at 140 °F for 4 h.

### 2.3. Four-Point Bend Experimental Testing

Experimental testing was conducted by Heng et al. [14,16] to investigate damage initiation and propagation and to generate validation data for the computational model development. The four-point bending tests were performed with an MTS Criterion 45 load frame that collected the loading force and displacement of the pins while the DIC system measured the strain field of each specimen. Specimens were cut from a layered structure panel using an Omax 2626 JetMachining Center. Figure 3a shows the testing configuration with the loading pins placed at one-third of that of the support pins. The MTS flexure grips were pins with a 0.2-inch diameter. The loading rate of the MTS load frame was set to 0.05 inch/min following ASTM standards. Both the MTS load frame and the DIC system collected data with the same frequency and were initiated at the same time to correlate the data. The front face of all specimens was sprayed with a flat white paint and then speckled with a flat black paint to create a pattern for use with the DIC (Figure 3b). The image correlation software VIC created subsets on the face of the specimen, and the black speckle displacements for each subset were analyzed during loading. The software then performed strain calculations to present strain fields for the speckled face of each specimen.

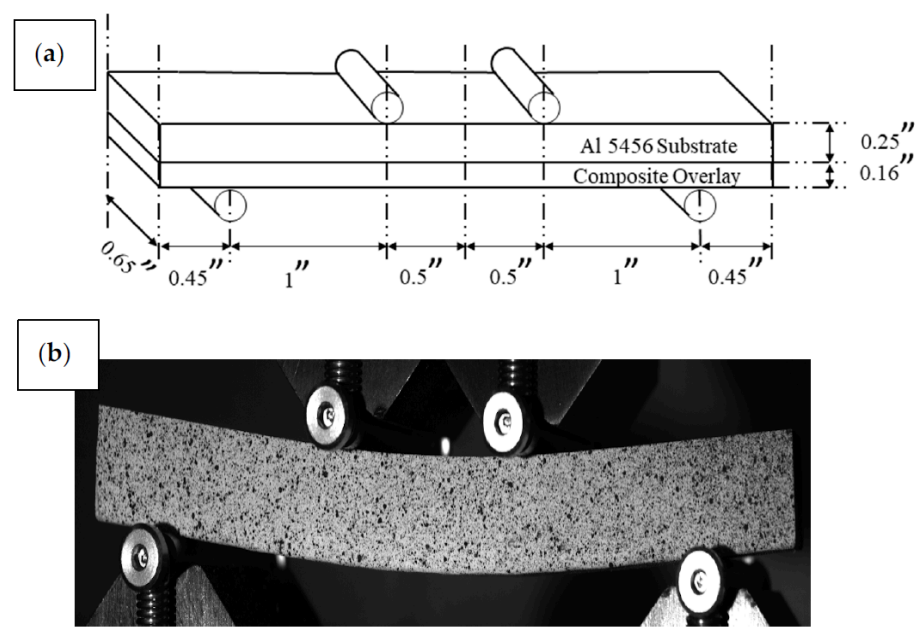

**Figure 3.** (**a**) Dimensions of experimental configuration. (**b**) Image of specimen during testing [14].

### 2.4. Material Properties

The material properties for the FE source model were acquired from multiple, disparate sources. Some experimental data were available from the Naval Surface Warfare Center Carderock Division [53] while other parameter data were approximated using literature values for similar materials or obtained directly from manufacturer data sheets. In many cases, data were sparse for a given parameter, and in some cases only a single average value was used to represent a material property. These 41 mechanical properties populated the parameter space evaluated in this study and full details can be found in the previous literature on the source model development [14-17]. Only material parameters were selected to limit the scope and to focus on investigating the accuracy and efficiency of the RDSM summation approach.

The aluminum 5456 substrate was modeled as an isotropic material with mechanical properties generated through cylindrical tensile testing as provided in Table 2 [53]. The Johnson–Cook equation (Equation (1)) was used to describe the strain hardening phenomenon as

$$\bar{\sigma} = A + B\bar{\varepsilon}^n \tag{1}$$

where $\bar{\sigma}$ is the equivalent plastic stress, $\bar{\varepsilon}$ is the equivalent plastic strain, and $A$, $B$, and $n$ are material constants. The aluminum mechanical properties constituted 5 of the 41 parameters in the model. The properties of the resin between the laminae and comprising the composite/metal interface are provided in Table 3. Double Cantilever Beam (DCB) and End Notched Flexure (ENF) tests on a laminate fabricated with Hexcel 7500 were performed to obtain the energy release rates for mode I and II following ASTM D5528 and D7905, respectively [53]. This set of properties was used twice in the source model, once for the cohesive layers between the lamina plies and again for the cohesive layer between the composite and the aluminum substrate, thus accounting for 12 of the 41 parameters.

**Table 2.** Elastic and plastic properties for aluminum 5456-H116.

| Mechanical Property | Symbol | Average Value |
|---|---|---|
| Young's Modulus | $E$ | 10.1 msi |
| Poisson's Ratio | $v$ | 0.29 |
| Yield Stress | $A$ | 29.8 ksi |
| Strength Coefficient | $B$ | 103.6 ksi |
| Strain Hardening Exponent | $n$ | 0.607 |

**Table 3.** Mechanical properties of the resin/hardener system M1002/M2046.

| Mechanical Property | Symbol | Average Value |
|---|---|---|
| Elastic Modulus | $EC$ | 10 msi |
| Nominal Stress Normal-only Mode | $XT$ | 7.6 ksi |
| Nominal Stress First/Second Direction | $XS$ | 4.9 ksi |
| Normal Mode Fracture Energy | $GI$ | 7.6 lb-in/ in$^2$ |
| Shear Mode Fracture Energy First/Second Direction | $GII$ | 16.6 lb-in/in$^2$ |
| Mixed Mode Behavior for Benzeggagh–Kenane | $BK$ | 2.6 |

The properties for the lamina plies are provided in Table 4 and Table 5. Lack of shear property data caused each of the four lamina types to be assigned the same shear properties. This assumption was considered acceptable as each of the four lamina types has the same matrix that governs the shear properties. The intralaminar fracture toughness of the Hexcel laminae were estimated from similar E-glass/epoxy weave fabric from Mandell et al. [54], while the intralaminar fracture toughness of the E-BX/E-LT laminae were estimated from Hallet et al. [55]. The Young's modulus and tensile strength of the E-BX/E-LT laminae were obtained directly from the manufacturer's data sheets. The shear properties were derived using a technique detailed by Johnson et al. [56]. Each lamina ply required 6 mechanical properties in the source model. With four variations of plies, 24 of the 41 parameters defined the composite plies. Additional information on data collection can be found in Heng's dissertation [16].

**Table 4.** Mechanical properties of lamina plies.

| Mechanical Property | E-BX 1200 | | E-LT 1800 | | Hexcel 7500 | | Hexcel 7781 | |
|---|---|---|---|---|---|---|---|---|
| | Symbol | Average Value | Symbol | Average Value | Symbol | Average Value | Symbol | Average Value |
| Young's Modulus | $E1200$ | 2.8 msi | $E1800$ | 2.8 msi | $E7500$ | 2.83 msi | $E7781$ | 4.4 msi |
| Tensile Strength | $X1200$ | 53 ksi | $X1800$ | 53 ksi | $X7500$ | 46.7 ksi | $X7781$ | 70 ksi |
| Poisson's Ratio | $V1200$ | 0.15 | $V1800$ | 0.15 | $V7500$ | 0.15 | $V7781$ | 0.15 |
| Intralaminar Fracture Toughness | $G1200$ | 150 lbs/in | $G1800$ | 150 lbs/in | $G7500$ | 100 lbs/in | $G7781$ | 100 lbs/in |

**Table 5.** Laminar shear properties of the lamina plies.

| Mechanical Property | Symbol | Average Value |
|---|---|---|
| Shear Modulus of Lamina | $GS$ | 0.8 msi |
| Shear Strength of Lamina | $SS$ | 5.16 ksi |
| Shear Damage Parameter | $alpha12$ | 0.2767 |
| Maximum Shear Damage | $d12$ | 0.714 |
| Maximum Shear Plastic Strain | $epsilon$ | 0.02 |
| Effective Shear Yield Stress | $sigmaY$ | 5.16 ksi |
| Coefficient in Shear Hardening Equation | $C$ | 0.65 msi |
| Power Term in Shear Hardening Equation | $P$ | 0.729 |

*2.5. Computational Model Development and Validation*

While manufacturing and testing layered structures to characterize the large parameter space is prohibitive, validated computational simulation provides an effective capability to rapidly explore the influence of the material properties on damage tolerance and to characterize the interactions of damage mechanisms throughout the parameter space. Structural analysis solutions exist for lap and scarf joints (e.g., [57,58]); however, advanced, high-fidelity analysis, such as FE analysis, is required to capture the effects of the multi-physics, interacting damage mechanisms for generic configurations of layered structures. A validated source model was developed by Heng et al. and Arndt et al. [14,15] that explicitly models each lamina layer of the composite overlay, the metallic substrate, and the composite/metal interface to predict structural damage tolerance (Figure 4). This high-fidelity 3D FE model for analysis with ABAQUS [59] captures the interacting damage mechanisms including matrix cracking (*PL*), fiber fracture (*DL*), delamination within the composite (*DC*), disbond at the composite/metal interface (*DI*), and yielding in the metal (*PM*).

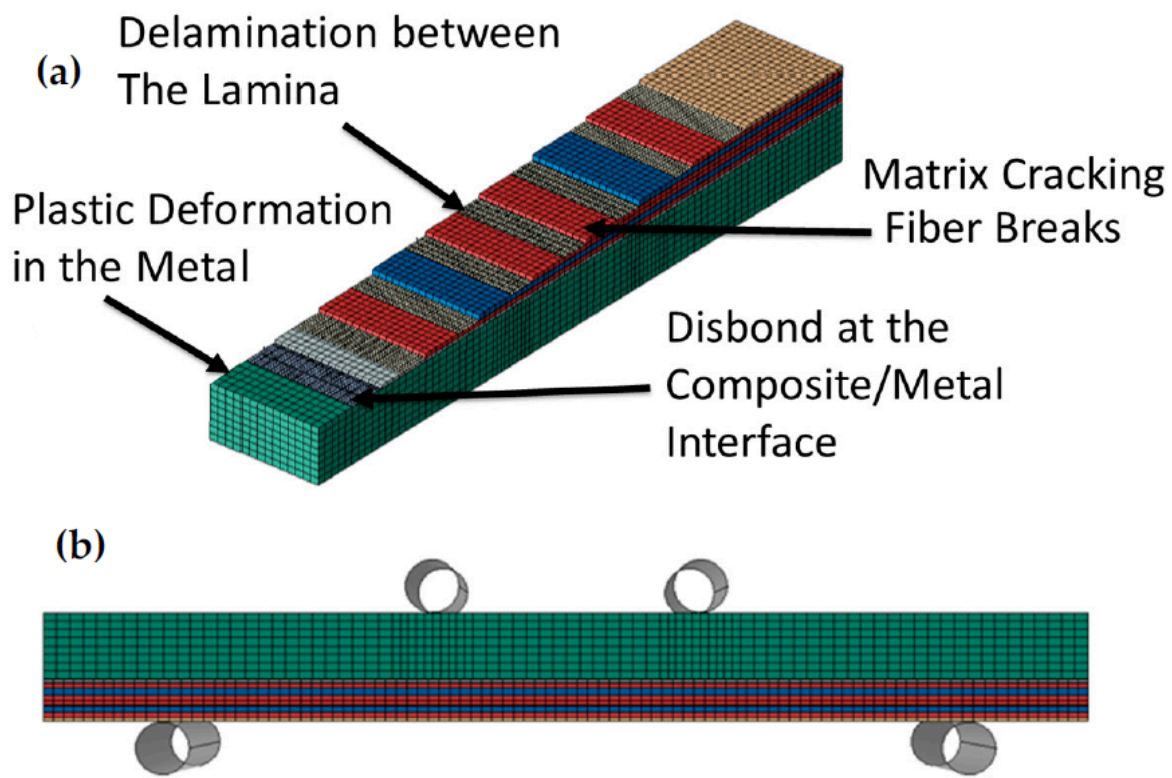

**Figure 4.** (**a**) The laminae, cohesive layers, interfaces, aluminum substrate, and corresponding damage mechanisms are modeled explicitly in the FE source model of the layered structure. (**b**) The source model under four-point bending with the aluminum on top.

Each of the fabric layers were modeled explicitly with continuum shell elements (SC8R, 3 degrees of freedom). The elements for all lamina layers were seeded at 0.04 inches. The continuum damage mechanics (CDM) damage model of each lamina is incorporated using a built-in ABAQUS VUMAT user subroutine [56,60]. Cohesive elements (COH3D8, 3 degrees of freedom) were included between each lamina and at the composite/metal interface to capture delamination and disbond, respectively. These elements utilized the triangular traction–separation law integrated in ABAQUS with the cohesive zone model (CZM). The element size, stiffness, and strength for these cohesive elements were selected according to guidelines from Turon et al. [61]. The cohesive elements were seeded at 0.02 inches. The aluminum 5456 substrate was modeled using solid elements (C3D8R, 3 degrees of freedom). This part was meshed with two distinct seeds. The global seed was set to 0.05 while the regions underneath the loading pins were seeded at 0.03 to transfer the loading properly. Rigid bodies were used to model the support and loading pins to represent the experimental boundary and loading conditions. The resulting model consisted of a total of 111,291 elements with a total degree of freedom of 664,407. To characterize the varying damage mechanisms and to automate output extraction for computational efficiency, the total damage absorption energy and mechanism energies were the output.

### 2.5.1. Damage Modeling

Damage modeling in the layered structure requires the interaction of different methodologies. The CDM is used to model the intralaminar damage of fiber fractures and matrix cracking. The CZM is used to model the interlaminar damage (delamination) and the disbond of the interface between the composite overlay and the aluminum. Plasticity is included as the only damage type in the aluminum. A preliminary FE investigation using the Johnson–Cook damage model indicated zero effect on results, and no damage other than plastic deformation was observed in the metal during experimental testing. Therefore, damage in the metal is not included in this study (although fracture in the metal is another damage mode that should be included in a general analysis of layered structures). These damage models are presented in detail by Heng et al. and Arndt et al. [14-17] and are summarized here.

CDM for Intralaminar Damage

The CDM approach has been extensively applied to predict composite failure modes and integrates stress and strain failure and fracture mechanics. Damage initiation is predicted by the stress and strain failure criteria, while the damage evolution is captured by fracture mechanics through damage variables. Microscale fiber and matrix cracks and the plasticity of the matrix gradually degrade the load carrying capacity of the composite prior to complete failure. The microcracks are quantified at the macro-scale by the CDM to degrade the material properties. Damage variables inform the progressive reduction in the material stiffness. Each lamina is modeled as a homogeneous orthotropic material. The fiber-dominated tensile or compressive damage is modeled by the elastic damage model. The matrix-controlled shear failure is modeled by the elastic-plastic damage model.

It is typical to apply the theory of effective stress from the strain equivalence theory to represent the constitutive equations for laminae with damage variables in the elastic domain using plane stress (Equation (2)).

$$\left[\varepsilon_{11} \; \varepsilon_{22} \; \varepsilon_{12}^e\right] = \begin{bmatrix} \frac{1}{E_{11}} & \frac{-\nu_{12}}{E_{11}} & 0 & \frac{-\nu_{21}}{E_{22}} & \frac{1}{E_{22}} & 0 & 0 & 0 & \frac{1}{2G_{12}} \end{bmatrix} \left[\overline{\sigma}_{11} \; \overline{\sigma}_{22} \; \overline{\sigma}_{12}\right], \; \left[\overline{\sigma}_{11} \; \overline{\sigma}_{22} \; \overline{\sigma}_{12}\right] = \left[\frac{\sigma_{11}}{(1-d_{11})}\right] \quad (2)$$

where $d_{11}$ and $d_{22}$ are damage variables responding to fiber fracture along the 11 and 22 directions and $d_{12}$ is the damage variable associated with matrix deterioration in shear deformation. Tensile and compressive fiber failures are defined with Equation (3),

$$d_{11} = d_{11}^t \frac{\langle\sigma_{11}\rangle}{|\sigma_{11}|} + d_{11}^c \frac{\langle-\sigma_{11}\rangle}{|\sigma_{11}|}, \; d_{22} = d_{22}^t \frac{\langle\sigma_{22}\rangle}{|\sigma_{22}|} + d_{22}^c \frac{\langle-\sigma_{22}\rangle}{|\sigma_{22}|} \quad (3)$$

$$\langle x \rangle = \{ 0, \; x < 0 \; x, \; x > 0$$

where fiber fracture under tensile or compressive loading is described by $d_{11}^t$ and $d_{22}^t$, respectively, which are components of $d_{11}$.

The nominal stress, $\sigma$, causes strain on the damaged material and defines the stress on the undamaged material known as effective stress, $\overline{\sigma}$. This effective stress is a direct approach in defining the damage initiation and damage evolution. Once the effective stress reaches the material strength, damage is initiated according to Equation (4),

$$\frac{\overline{\sigma}_{ij}}{X_{ij}} = 1, \qquad i, j = 1, 2 \quad (4)$$

where the tensile and compressive strengths for the uniaxial loading along the fiber are defined with $X_{11}$ and $X_{22}$. $X_{12}$ is the shear strength. Initiation of the damage dictates the change in the evolution of $d_{11}$ and $d_{22}$ described by exponential equations, Equation (5),

$$d_{ii} = 1 - \frac{1}{k_{ii}} \exp\left\{-\frac{2U_0^{ii}L_c}{G_{ii}^f - U_0^{ii}L_c}(k_{ii} - 1)\right\}, \qquad i = 1, 2 \quad (5)$$

$$k_{ii} = \frac{\overline{\sigma}_{ii}}{X_{ii}}, \quad U_0^{ii} = \frac{X_{ii}^2}{2E_{ii}}$$

where $G_{ii}^f$ is the fracture energy of laminates in the ii direction, $U_0^{ii}$ is the elastic energy density when damage is initiated, and $L_c$ is the characteristic length of the element. Nondecreasing behavior of $d_{11}$ and $d_{22}$ is ensured by having $L_c$ satisfy the requirement described by Equation (6).

$$G_{ii}^f - U_0^{ii} L_c > 0 \tag{6}$$

Therefore, the element size needs to be small enough to meet this requirement when implementing the damage evolution in FE (Equation (7)),

$$d_{12} = \alpha_{12} \ln (k_{12}), \quad k_{12} = \frac{\overline{\sigma}_{12}}{X_{12}} \tag{7}$$

where $\alpha$ is a material constant measured by experimental testing. In addition to the elastic response, plasticity behavior should be included for fabric laminates under shear loading. This behavior is described by the Ludwik–Hollomon equation [62], as Equation (8),

$$\overline{\sigma}_{12} = \widetilde{\sigma}_y + C(\varepsilon_{12}^p)^P, \qquad \varepsilon_{12}^p = \varepsilon_{12} - \varepsilon_{12}^e \tag{8}$$

where $\widetilde{\sigma}_y$ is the effective stress corresponding to the normal stress at the yield point. $\varepsilon_{12}^p$ is the plastic part of the total strain $\varepsilon_{12}$. $C$ and $P$ are material properties of the matrix. When the plastic strain reaches a maximum value, $\varepsilon_{max}^p$, the lamina fails.

### CZM for Delamination and Disbond

CZM is applied to model the delamination between plies and the disbond at the interface. The triangular traction–separation law (Equation (9)) is selected for its widespread use [63].

$$\{t = K\delta, \qquad t \le t^0 \ t = \frac{t^0(\delta - \delta^f)}{\delta^0 - \delta^f}, \qquad t > t^0 \tag{9}$$

where $t$ is the traction, $K$ is the interface stiffness relating the traction and corresponding separation before the damage initiation, $\delta$ is the separation, $t^0$ is the damage initiation stress, $\delta^f$ is the maximum separation when the element fails, and $\delta^0$ is the separation at damage initiation. $G_c$ is the mixed-mode critical fracture energy and is the area under the traction–separation curve. Mixed-mode fracture initiation is predicted by the quadratic stress criterion, which determines damage initiation and evolution, Equation (10),

$$\left\{\frac{\langle t_{III} \rangle}{t_{III}^0}\right\}^2 + \left\{\frac{t_{II}}{t_{II}^0}\right\}^2 + \left\{\frac{t_I}{t_I^0}\right\}^2 = 1 \tag{10}$$

where $t_I$, $t_{II}$, and $t_{III}$ are tractions for Mode I, Mode II, and Mode III fractures. $t_I^0$, $t_{II}^0$, and $t_{III}^0$ are the damage initiation stresses for the three modes of fracture. When the left side of Equation (10) equals unity, damage is initiated. The mixed-mode fracture criterion is defined by one of the most widely used expressions for the critical release rate of mixed-mode loading, and was proposed by Benzeggagh and Kenane [64] (Equation (11)).

$$G_c = G_{Ic} + \left(G_{IIc} - G_{Ic}\right)\left(\frac{G_{II} + G_{III}}{G_I + G_{II} + G_{III}}\right)^n \tag{11}$$

$G_{IC}$ and $G_{IIC}$ are the critical energy release rates for Mode I and Mode II fractures. $G_I$, $G_{II}$, and $G_{III}$ are the energy release rates for the three modes of fracture. $n$ is a material property measured by experimental testing.

### 2.5.2. Model Validation

Validation for the FE model was assessed by two comparisons with the experimental results. The first was a comparison of force–displacement curves (Figure 5). The FE results reasonably align with that of the experimental results and capture the progressive damage of the overlay. The first peak corresponds to the intralaminar fracture and delamination of the first ply (Hexcel 7781). The shear failure of the fourth and fifth plies (E-BX 1200) caused the drop after the second peak. This observation indicates that the distance between the first two peak loads is controlled by the shear properties of the laminae. Some discrepancy is expressed by the graph in the elastic region and the second peak prior to the E-BX 1200 failure. This stems from several material properties characterized by sparse data. The overall correlation of the FE model was deemed acceptable for the application of reduced-dimension surrogate modeling. The results of the RDSM formulation process identify the most influential parameters, thus informing limited testing to fully characterize these properties.

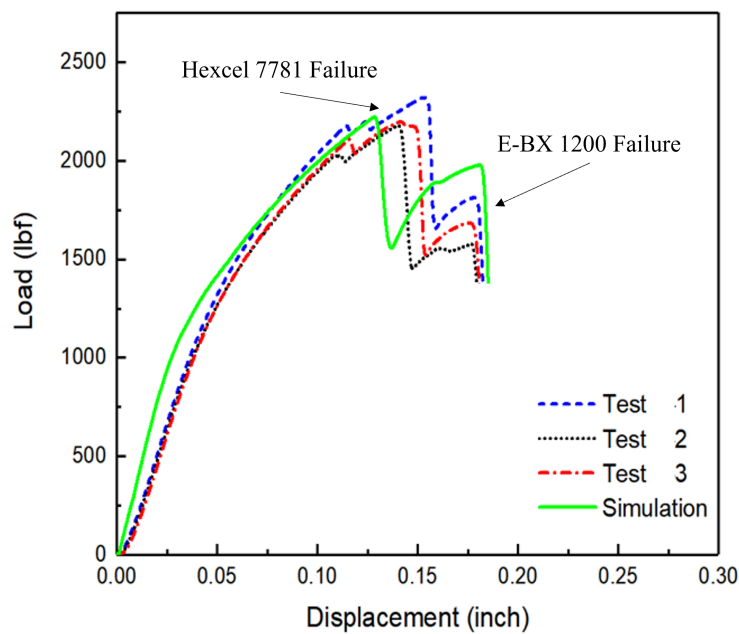

**Figure 5.** Comparison of force–displacement curves [14].

The second comparison evaluated the DIC strain fields relative to those generated with the FE model. Figure 6 illustrates that the FE model produced similar strain fields and locations of damage initiation and progression. Once validated, FE simulation was used to generate the damage energy output for varying parameter values. In the comparison, Lagrange strain was utilized as a built-in function of the DIC post-processing software. The Lagrange strain is calculated relative to the undeformed coordinate system of the specimen. High strain values (red) were observed underneath the loading pins and in the region of the final damage propagation where parts of the composite overlay were still supporting load.

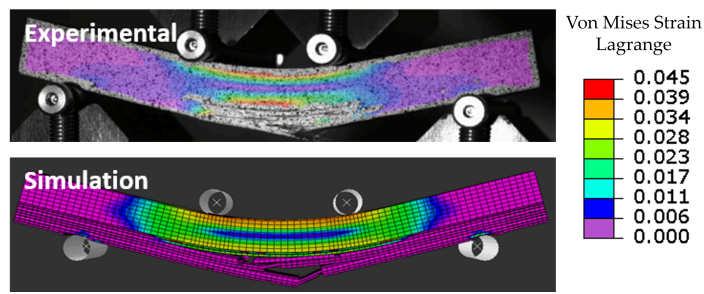

**Figure 6.** Experimental and computational strain fields at final failure.

2.5.3. Data Sampling

Populating a dataset using high-fidelity physics-based models can be very expensive computationally. For this reason, it is essential to efficiently sample the input space to minimize the number of model simulations while capturing the statistical properties of the response. Parameter sampling of the 41 mechanical property inputs was performed with Distribution-based Input for Computational Evaluations (DICE) [65]. DICE is a general-use software tool that generates ready-to-execute input files using parameter distributions. Due to data sparsity for the majority of the input parameters, Monte Carlo sampling was performed using a uniform distribution with of range of ±20% of the mean values for surrogate modeling and sensitivity analysis. A normal distribution with the standard deviation set to 10% of the mean value was applied for uncertainty quantification (UQ) to determine if an adequate number of parameters were included in a RDSM. The surrogate models were sampled for sensitivity analysis using Latin Hypercube Sampling (LHS) [66], one of the most common sampling methods, while Latin Stratified Sampling (LSS) was applied for UQ [67].

## 2.6. Reduced-Dimension Surrogate Modeling

Both paths for RDSM formulation are based on surrogate modeling and sensitivity analysis, although these methods are performed at different stages (Figure 2). RDSMs are generated in the form of surrogate models with a reduced number of parameters. The parameter space is initially reduced through the calculation of false discovery rate (FDR) measures.

### 2.6.1. Surrogate Modeling

Surrogate models are analytical approximations that can predict the behavior of high-fidelity FE models using a limited number of outcomes determined from adequately sampled input parameter sets. One of the most widely used surrogate modeling techniques, artificial neural network (ANN), was selected to generate the surrogate models [68], where an approximation function $\hat{f}$ represents the true model $f$. These approximate models replicate the behavior of the complex high-fidelity model with a fraction of the computational cost. With ANN, the approximation function $\hat{f}$ describes a set of features organized with a hierarchy. Each layer starts by taking the previous layer's inputs and generates the output for the next layer. Each layer consists of nodes, or neurons, that represent a unit of computation in the approximation. The nodes perform a linear transformation of one or more inputs (Equation (12)) followed by a nonlinear transformation applied by the activator ($\sigma(\cdot)$).

$$z = \sum_i^H w_i \bullet x_i + b \tag{12}$$

$z$ is the weighted sum of the inputs, $x_i$ are the inputs to the node, and $w_i$ is the weight coefficient for the linear transformation. These weight coefficients are tunable and are adjusted to minimize the error between each layer. The activator function is then applied to this weighted sum before being passed to the next layer or output.

The surrogate model for *TS* was generated using Google TensorFlow, an open-source software library [69]. A total of 1555 data points were generated from the DICE sampling of the 41 input parameters and the corresponding output energies were acquired through FE analysis. The *TS* surrogate model was generated with all 41 input parameters and the total damage energy absorption as the output. Each of the mechanism surrogate models was generated on reduced-input parameter spaces as determined by FDR logworth parameter screening using the specific mechanism energy as the output. This process is described in Section 3.2 along with the results from the FDR logworth screening. The ANN consisted of a Sequential Keras model implementing a Scikit-Learn KerasRegressor interface with an Adam optimizer, a normal initializer, and a ReLU activator. The number of hidden layers and nodes per layer were swept until an adequate fit was established while restricting the overall structure of the ANN to a reasonable number of layers and node structure to avoid overfitting of the output.

### 2.6.2. Sensitivity Analysis

While surrogate models provide fast-running formulations to predict outcomes, surrogate model accuracy is highly dependent on the number of data points generated from the source model (correlating the input parameters to an output energy) and the sampling method that identifies the parameter values used to generate the output. Accuracy will be compromised if there are not enough data points representing highly influential parameters and parameter subspaces as well as parameter subspaces with rapidly varying behavior. Therefore, complementing surrogate model formulation with sensitivity analysis can limit the parameter space for improved efficiency. Sensitivity analysis is a method used to identify the most influential parameters on the output. Then, only the most influential parameters are used as variables in surrogate model formulation while all other parameters are held constant at average or design values to create a RDSM. Sajedi and Liang implemented a similar approach utilizing surrogate model generation to classify the controlling damage mechanisms for an overall concrete structure [70].

#### FDR logworth

As global sensitivity analysis (GSA) is computationally intensive for high-dimensional, complex problems, FDR logworth [71,72] was explored as a screening tool to reduce the parameter space prior to GSA. FDR logworth typically requires significantly less computational time than GSA, is easy to perform, and is available in widely used statistical analysis tools such as SAS [73]. As FDR logworth only calculates total influence measurements, it does not provide the parameter interaction information available through the more advanced GSA methods.

FDR is a statistical method for conceptualizing the rate of type I (false positive) errors in the null hypothesis and is conducted with multiple comparisons [74]. A *p*-value < 0.05 generally signifies that a parameter is significant; however, for the layered structure problem, the p-values of the dominant parameters were extremely low and difficult to differentiate. Therefore, FDR logworth measures were calculated. In an FDR logworth analysis, the logworth of the FDR p-value is calculated (Equation (13)).

$$FDR(LogWorth) = -\log_{10}(FDR\ p - value)$$

(13)

With FDR logWorth, the higher the value, the more significant the parameter. To evaluate FDR logWorth as a screening tool, FDR logworth and GSA results were compared at various stages of RDSM formulation and agreed in all instances.

#### Global Sensitivity Analysis

GSA was performed by calculating Sobol' indices [75]. This method is a commonly used variable-based approach that measures each parameter's influence by delineating its contribution to the total variance of the output. The first-order effect measures the influence of an individual input parameter alone, and the total-order effect includes the interactions of the parameter under consideration with all other parameters. Comparison of the first-order and total-order indices for a given parameter provides insight into parameter interactions. Similar indices indicate little or no interaction with other input parameters, while a large difference indicates significant interaction. The Sobol' indices method also allows for the investigation of higher-order effects, the contributions of a specific set of variables together, to further investigate parameter interactions. Sobol' indices evaluation requires a heavily populated dataset to converge which was implausible to generate using the FE source model; therefore, a surrogate model was necessary to perform this type of GSA.

Sobol' indices are calculated according to Equation (14).

$$Y = f(X), \ X = \{X_1, X_2, X_3, \ldots X_m\} \tag{14}$$

where $X$ is the vector of $m$ inputs, $Y$ is the model output, and $f$ is a square integrable function. The function is expanded according to Equation (15).

$$(X) = f_0 + \sum_{i=1}^{m} f_i(X_i) + \sum_{i<j}^{m} f_{ij}(X_i, X_j) + \ldots + f_{1,2,\ldots,m}(X_1, X_2, \ldots, X_m) \tag{15}$$

in which $f_0$ is a constant, and the other terms are functions of corresponding inputs with a zero mean. Squaring both sides of Equation (15) results in Equation (16).

$$V(Y) = \sum_{i=1}^{m} V_i + \sum_{i<j}^{m} V_{ij} + \ldots + V_{1,2,\ldots,m} \tag{16}$$

where $V(Y)$ is the variance of $Y$, $V_i$ is the variance of $f_i(X_i)$, and so on. Dividing both sides of Equation (16) by $V(Y)$ yields Equation (17).

$$\sum_{i=1}^{m} S_i + \sum_{i<j}^{m} S_{ij} + \ldots + S_{1,2,\ldots,m} = 1 \tag{17}$$

where $S_i$ is the first-order index, $S_{ij}$ is the second-order index, and so on. $S_i$ is the influence of $X_i$ on the variance of the output and $S_{ij}$ is the influence of interactions between $X_i$ and $X_j$ on the variance of the output. Higher-order effects were not part of this investigation; as such, only the first-order and total-order indices were calculated.

$S_i$ for a single parameter is the ratio of two variance computations, as in Equation (18). The numerator is calculated as the variance of the mean of the output parameter conditioned on a single fixed input parameter. The mean in the numerator is evaluated numerous times across the sample space, thereby incorporating the impact across the entire space of the input parameter range. The denominator normalizes the index with respect to the variance across the entire sample space, allowing for comparison across various input parameters for the same output parameter. An input parameter with a larger Sobol' index is more impactful across the sample space than other input parameters for that same space with lesser Sobol' indices.

$$S_1^i = \frac{V_{x_i}(E_{x_{\sim i}}(X^i))}{V(Y)} \tag{18}$$

In addition to evaluating the first-order effects of a single input parameter, total effects are also commonly reported for GSA results. The total effects of a parameter attempt to identify not only the independent parameter effects but also effects due to all interactions with the parameter under investigation. While the calculation to independently measure the Sobol' index for all higher-order effects can be performed to determine this total-effects index, Equation (19) demonstrates the commonly utilized analog which evaluates the total effects as unity minus the effects of all parameters except those under evaluation. This alteration provides a significant improvement to the computational efficiency of the process.

$$S_{Ti} = 1 - \frac{V[E(X_{\sim i})]}{V(Y)} \tag{19}$$

$E(Y | X_{\sim i})$ are the conditional expectations of the output $Y$ when input $X_i$ is not considered. It should be noted that all parameters investigated in the current effort are considered as independent parameters. Correlation was not included in GSA and should be accounted for in future work.

## 3. Results

The results are presented following the two pathways in Figure 2. The first path provides the results for the RDSM determined directly from the total damage energy output. The second path presents the mechanism RDSM formulations and characterization performed using the individual damage mechanism energy outputs. A third section compares the two methods of RDSM formulation. RDSM formulation for both pathways was based on the 1555 FE results obtained from the source model and sampled with all 41 parameters. As shown in Figure 7, the total energy varies throughout the parameter space as do the contributions of the damage mechanisms. It should be noted that *DI* contributed up to 7.6% for some parameter combinations comprising a small subspace of the entire parameter space and confirming the importance of characterizing the damage mechanisms separately to ensure adequate sampling in small but significant subspaces. (A subspace is defined as a distinct region within the full parameter space.) Overall, *PM* dominated the parameter space. The next two highest energy contributors were *PL* and *DC*.

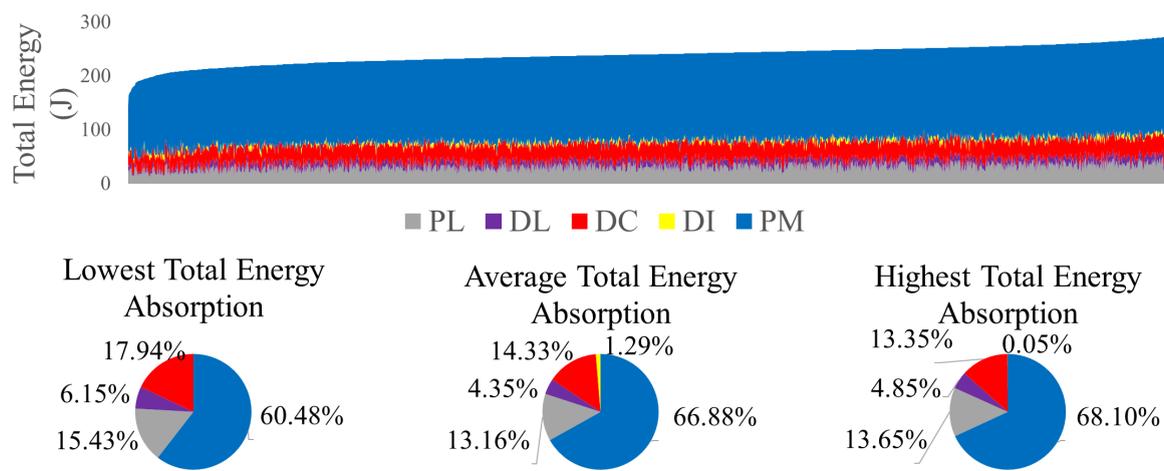

**Figure 7.** Distribution of energies absorbed by the FE model detailing the contribution of various damage mechanisms.

### 3.1. Direct RDSM Formulation

A surrogate model with all 41 input parameters as variables was generated using the FE source model dataset. The output layer consisted of a single output, the *TS* energy absorption. The dataset consisted of 1555 data points, and 25 of these points were randomly selected to create a validation set. The remaining 1535 data points were split into testing and training data with a 10/90 split. KerasRegressor with two hidden layers of 60 and 80 nodes required the use of an Adam optimizer, a normal initializer, and a ReLU activator. The learning rate was set to 0.001. The surrogate model resulted in a mean absolute error (MAE) of 3.45%. The accuracy of the resulting *TS* surrogate model is illustrated in Figure 8, which presents a 45° plot of the 25 randomly saved validation data points. The plot shows the predicted results from the *TS* surrogate model versus the values calculated by the FE source model. Predictions fall within 5% error of the actual with only three data points outside of this bound. These three outliers fell within 10% of the calculated FE results. Given the complexity of the parameter space and the limited data sampling, this accuracy was considered adequate to avoid overfitting the model with additional training. It should be highlighted that development of the surrogate model was user- and time-intensive due to the number of parameters.

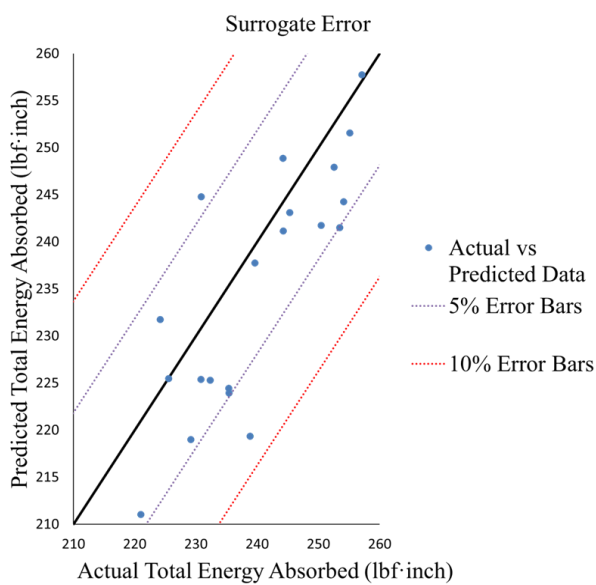

**Figure 8.** Error of 25 randomly selected data points generated from the *TS* surrogate model.

GSA to calculate Sobol' indices was performed using the *TS* surrogate model to inform parameter space reduction. The first-order and total-order Sobol' indices of all 41 input parameters were calculated with regards to the *TS* parameter space (Figure 9). Adequate convergence was achieved with 10,000 samples as results with 200,000 samples exhibited no deviations in the controlling parameters or their minimal interactions. The top four most influential parameters were *A*, *E*, *XS*, and *Aln*. Three of the four most influential parameters (*A*, *E*, and *Aln*) represent *PM*, the most dominant damage mechanism. The strength coefficient, *B*, which is the other parameter representing *PM*, is within the eight most influential parameters. Shear properties of the resin, *XS*, *P*, and *GII* were in the top six most influential parameters. Minimal interaction was indicated between the dominant parameters when comparing first-order and total-order indices. Large drops in the index values are evident after the first- and fourth-highest contributors.

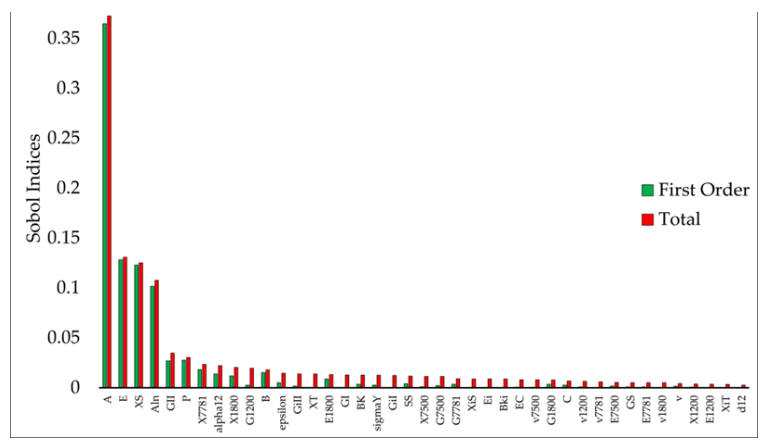

**Figure 9.** Sobol indices for the *TS* parameter space.

Sobol' indices' calculation was manually intensive and time-consuming. Therefore, FDR logworth was performed to explore its potential as a screening tool to reduce the parameter space prior to surrogate modeling and GSA. Table 6 compares the total-order index from Sobol' indices to the FDR logworth values. Both methods identically ordered the most influential parameters and captured the drops after the first and fourth dominant parameters. FDR logworth provided comparable information regarding the importance of the parameters at a significantly reduced computational cost. As such, FDR logworth was employed for parameter space reduction prior to mechanism RDSM formulation.

**Table 6.** Total-Order Indices vs. FDR Logworth values of the *TS* parameter space.

| Variable | Total-Order Index | FDR Logworth |
|----------|-------------------|--------------|
| A | 0.37 | 59.38 |
| E | 0.13 | 18.67 |
| XS | 0.11 | 12.99 |
| Aln | 0.11 | 12.54 |
| GII | 0.03 | 3.96 |
| P | 0.03 | 3.90 |

UQ was conducted on the most influential parameters to determine which parameters to include in the RDSM. Beginning with only the most dominant parameter, *A*, the surrogate model was queried by varying only the parameter(s) under consideration while the remaining parameters were held constant at their mean values. Next, the top two parameters were varied and so on until minimal error with the inclusion of the next parameter as a variable was achieved. Convergence dictated that 5000 data points were needed for UQ. Negligible change in the mean *TS* prediction was observed when only the most dominant parameter *A* was included in the RDSM (Table 7). As the standard deviation exhibited notable improvements until the addition of the fifth most influential parameter, the top four parameters were included in the *TS* RDSM which was generated as part of the UQ.

**Table 7.** *TS* prediction results using RDSMs formulated as a function of the most influential parameters.

| | Total Energy Absorbed, *TS* (lbf-inch) | |
|----------|-----------|-----------|
| | **Mean** | **Std** |
| A | 234.807 | 5.843 |
| % Difference | 0.009% | 14.901% |
| A, E | 234.829 | 6.784 |
| % Difference | 0.001% | 8.044% |
| A, E, XS | 234.832 | 7.353 |
| % Difference | 0.035% | 5.521% |
| A, E, XS, Aln | 234.914 | 7.770 |
| % Difference | 0.003% | 1.654% |
| A, E, XS, Aln, P | 234.921 | 7.900 |

The RDSM formulated with the top four influential parameters as variables was used to map the *TS* parameter space (Figure 10). The 2D plots along the diagonal map each of the four parameters against the *TS* damage energy. As predicted by GSA, minimal interaction is indicated by the near linear gradations in all of the plots.

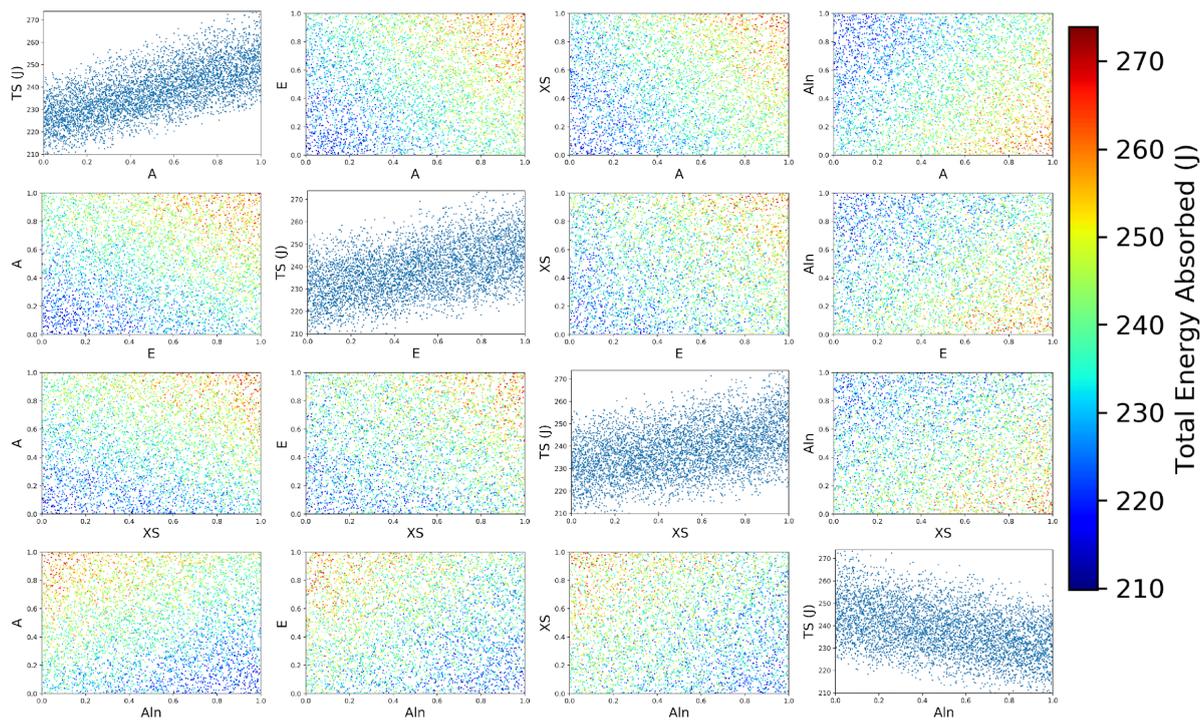

**Figure 10.** Four-dimensional *TS* parameter space visualization contoured with the total energy absorbed.

### 3.2. Mechanism RDSM Formulation and Characterization

Based on the previous results, FDR logworth was used as a screening tool to reduce the number of parameters prior to formulation of the surrogate models for each damage mechanism. This parameter reduction enabled an initial formulation of the mechanism RDSMs based on a reduced parameter set, eliminating the need to create a surrogate model based on all 41 parameters. The FDR logwoth measures for each damage mechanism are provided in Table 8. An FDR logworth value of 1.3 (*p*-value of 0.05) was considered potentially significant and included in the initial RDSM formulation. As seen in Table 8, there are considerable drop-offs in the FDR logworth values after the first few parameters for each damage mechanism. Additional investigation demonstrated that only the parameters before this drop significantly influenced RDSM accuracy. As a result, at most three parameters per damage mechanism were included in the mechanism RDSM formulations. This substantial reduction in variables based on FDR logworth screening decreased RDSM formulation time to the extent that this approach could be readily and rapidly applied to existing datasets. The reduced parameter space for each of the damage mechanisms is highlighted in Table 8.

**Table 8.** FDR logworth values for the subspace damage mechanisms.

| PM | | PL | | DL | | DC | | DI | |
|---|---|---|---|---|---|---|---|---|---|
| **Parameter** | **Value** | **Parameter** | **Value** | **Parameter** | **Value** | **Parameter** | **Value** | **Parameter** | **Value** |
| A | 100.56 | X1800 | 42.18 | XS | 25.97 | E1800 | 34.86 | P | 42.18 |
| E | 45.35 | XS | 38.01 | X7781 | 18.08 | X1800 | 14.12 | X7500 | 9.04 |
| Aln | 18.08 | E1800 | 24.83 | E1800 | 11.00 | P | 6.68 | epsilon | 8.25 |
| P | 6.30 | P | 8.12 | alpha12 | 7.79 | alpha12 | 5.36 | XiS | 6.83 |
| B | 3.34 | GII | 5.14 | P | 4.74 | G1800 | 4.55 | GS | 5.93 |
| epsilon | 1.86 | epsilon | 2.74 | GII | 2.40 | X7781 | 4.49 | GiII | 3.35 |
| C | 1.73 | G7781 | 2.57 | E7781 | 1.75 | X1200 | | C | 2.91 |
| SS | 1.33 | C | 2.43 | SS | 1.43 | SS | 3.91 | XS | 2.67 |
| | | sigmaY | 2.11 | G1800 | 1.43 | E1200 | 2.20 | E7500 | 2.60 |
| | | E1200 | 1.94 | | | XS | 2.18 | GII | 2.18 |
| | | X7500 | 1.37 | | | A | 1.96 | E1200 | 1.80 |

In general, mechanism RDSMs exhibited higher MAEs than the *TS* RDSM (Table 9). Parameter sampling was uniform across the entire parameter space; however, individual mechanism contributions and their gradients varied significantly throughout the parameter space (as will be demonstrated). For improved RDSM accuracy, adequate data are needed within each mechanism subspace (the region within the parameter space where a damage mechanism significantly contributes). This issue was highlighted in the initial attempt at RDSM formulation for the *DI* subspace, which comprised a much smaller region within the parameter space than the other mechanisms. As will be discussed, the *DI* damage energy was zero or near zero within most of the parameter space, with a very limited subspace in which *DI* significantly contributed to the total damage energy. This scenario hindered surrogate model generation, even with a reduced number of parameters, as the formulation process skewed to the data with zero contribution. Thus, the ANN fit the surrogate model to the much larger non-contributing region of the parameter space, thus inadequately capturing *DI* contributions within the subspace of significant damage energy. As such, it was not possible to formulate a RDSM with the available dataset. Additional sampling and FE analysis with the source model using only the most influential *DI* parameters according to the FDR logworth results was required to provide adequate output data in the *DI* subspace to then formulate a RDSM.

**Table 9.** RDSM prediction errors for the damage mechanisms.

| PL | | DL | | DC | | PM | |
|---|---|---|---|---|---|---|---|
| **Layers/Nodes** | **MAE** | **Layers/Nodes** | **MAE** | **Layers/Nodes** | **MAE** | **Layers/Nodes** | **MAE** |
| 2/65–70 | 6.60% | 2/55–55 | 14.45% | 1/50 | 4.99% | 2/60–80 | 2.29% |

The mechanism RDSMs were constructed with the same surrogate modeling process as the *TS* model, as detailed in Section 3.1. The primary difference in the mechanism RDSM formulation was that parameter reduction was performed prior to surrogate modeling. The *PM*, *PL*, and *DL* parameter spaces were reduced to three input parameters and one output, while the *DC* parameter space was reduced to two input parameters and one output. The MAE for each of the mechanism RDSMs is provided in Table 9. The *PM* RDSM exhibited the lowest error, because this damage mechanism was dominant across the entire parameter space, and thus more uniformly sampled than the other mechanisms. The collection of additional data to improve the *DL* RDSM was discussed; however, it was not performed as the summed RDSM error was comparable to that of the *TS* RDSM.

To characterize the damage mechanism behaviors, Sobol' indices were calculated using data generated from the mechanism RDSMs. Figure 11 shows parameter interactions for all mechanisms except PM. Figure 11a for *PL* illustrates some interactions between *E1800*, *XS*, and *X1800*, and Figure 11b shows interaction between *XS* and *E1800* in the *DL* subspace, while *X7781* is independent. Given that *XS* is dominant in two subspaces, it is not surprising that it was one of the top four influential parameters in *TS*. The *DC* subspace (Figure 11c) shows that the two dominant parameters, *E1800* and *X1800*, are interacting and that *E1800* has a significantly stronger influence on the delamination. Lastly, the *PM* subspace (Figure 11d) shows no parameter interactions with *A* as the most dominant parameter followed by *E* and *Aln*, consistent with the GSA results for *TS*.

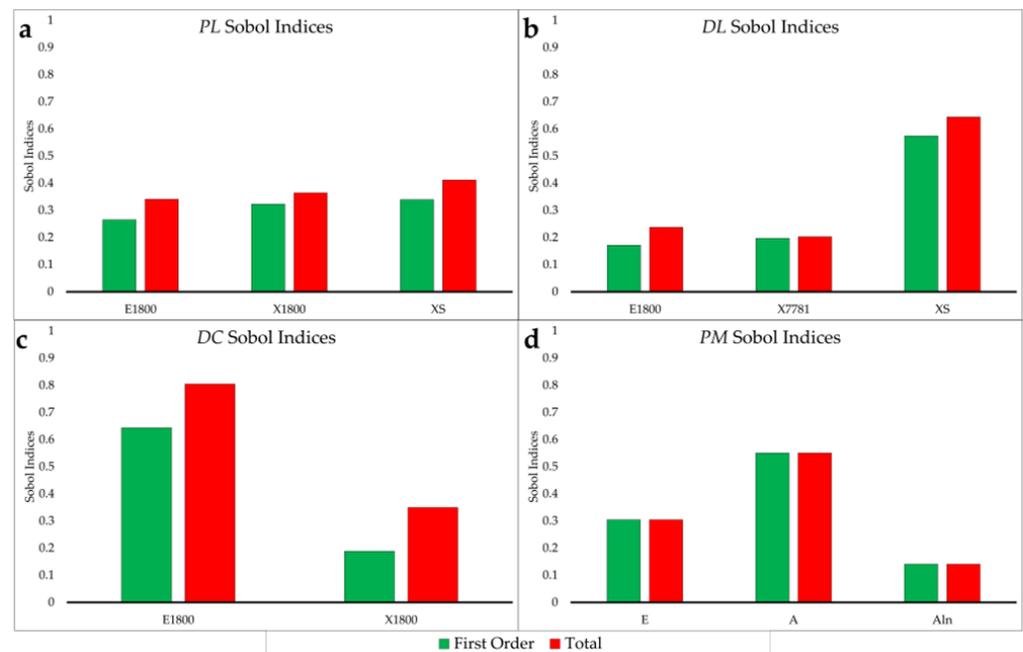

**Figure 11.** First and total indices for damage mechanisms. (**a**) Sobol' indices for shear plasticity in the laminate, *PL*. (**b**) Sobol' indices for intralaminar fracture in the laminate, *DL*. (**c**) Sobol' indices for interlaminar delamination, *DC*. (**d**) Sobol' indices for the plastic deformation in the metal substrate, *PM*.

The 3D contour graphs were generated for each mechanism to identify subspace locations and to understand parameter interactions (Figure 12 and Figure 13). Each mechanism RDSM was used to generate a 3D mesh grid consisting of 9261 datasets to uniformly and systematically capture the mechanism subspace. If the mechanism had a third dominant parameter, it was set at its minimum value when comparing the two other parameters.

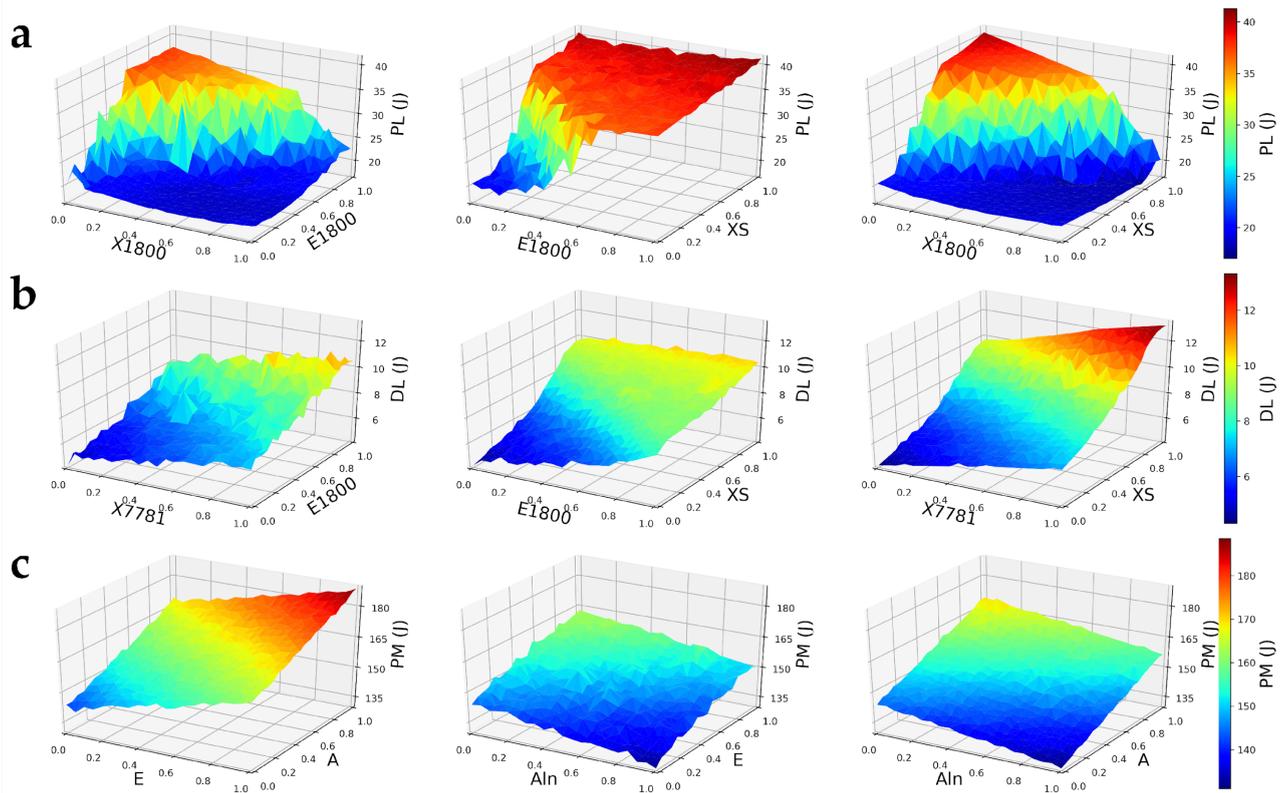

**Figure 12.** The 3D mesh contours for the damage mechanisms detailing potential interactions between dominant parameters. (**a**) Intralaminar fracture in the composite, *PL*. (**b**) Shear plasticity in the composite laminate, *DL*. (**c**) Plastic deformation in the aluminum substrate, *PM*.

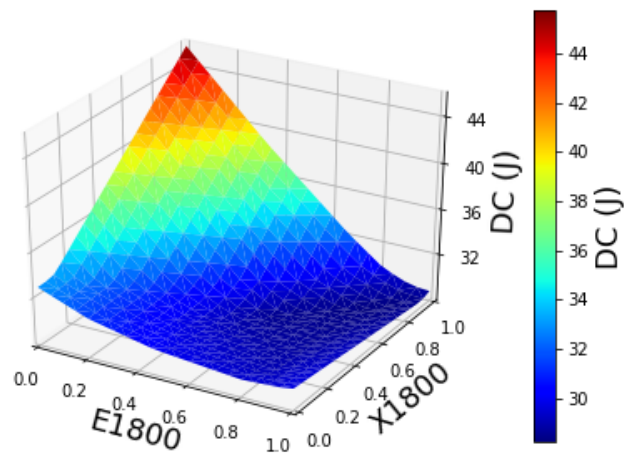

**Figure 13.** The 3D mesh contour for the intralaminar delamination between composite plies in the laminate, *DC*.

*PL* (Figure 12a) exhibits two distinct regions of low and high shear plasticity energy. The lower region is dominated by low values of *E1800* and *XS*. The higher energy region corresponds to high values of *E1800* and *XS* and low values of *X1800*. *DL* (Figure 12b) highlights that the intralaminar fracture of the composite overlay is controlled by the shear strength of the resin matrix, *XS*. This is noted by the large change in energy along its axis compared to the other two parameters, *X7781* and *E1800*. A higher *XS* requires a larger amount of energy for the damage to propagate through the laminae layers. This is mirrored with the tensile strength of the Hexcel 7781 layer, the outside ply, and *E1800* corresponding to the remaining interior layers carrying the stress after the outer plies fail. Figure 12c for the *PM* subspace illustrates that *A* is the dominant parameter with no interaction between the parameters. The metal substrate absorbs more energy when it is a stiffer less formable metal, providing it does not fracture. Increasing the yield stress and Young's Modulus and decreasing the metal's formability (*Aln*) will result in less deformation, allowing for the metal to absorb more energy before the outer composite overlay plies fail. Figure 13 shows that in the *DC* subspace, the delamination between the plies is controlled by the interior E-LT 1800 plies. As the stiffness of the plies is decreased, more deformation occurs leading to the resin carrying a larger stress. Increased delamination occurred with low *E1800* and high *X1800*.

*DI* Subspace Reduced-Dimension Surrogate Modeling and Characterization

As previously stated, the initial RDSM formulation for the *DI* subspace was not successful. Given the high distribution of the data as zero damage contribution, further FDR logworth was performed (Table 10) to separate the engaged (contributes 3% or more to total damage) and nonengaged (zero or <3%) regions. From these results, 12 parameters from the engaged group were selected for additional data generation to better sample the subspace where *DI* is active. While only three parameters were located above the drop applied previously, given the uncertainty in identifying the engaged subspace, the additional nine parameters were included in the sampling. Data acquisition followed the same procedure as that of the original data generation, except that only the 12 selected parameters were varied, while the remainder of the variables were held constant at their mean values. To expedite the data collection process, HPC was used to generate 3277 data points. The explicit FE source model required around 120 min per analysis to run locally but was reduced to 38 min on HPC systems. The dataset from the new sampling still resulted in many zero or near-zero outputs that continued to challenge RDSM formulation. Therefore, the 3% limit was again applied (8 joules of *DI*) and only the data above this value were included in the RDSM formulation (793 data points). The initial RDSM consisted of a surrogate model following the same framework presented in Section 3.1. For this RDSM, the 41 input parameters were reduced to 12 input parameters, and the output consisted of one variable (*DI*). The dataset consisted of 793 data points and utilized a 20/80 split for testing and training. A learning rate of 0.0015 was set for the model. A resulting feedforward surrogate with two layers of 16 neurons resulted in a MAE of 8.45%.

**Table 10.** FDR logworth values for the DI subspace reflecting engaged and nonengaged regions.

| Engaged | | Nonengaged | |
|---|---|---|---|
| **Parameter** | **Value** | **Parameter** | **Value** |
| XS | 41.81 | X7500 | 1.99 |
| P | 24.80 | XS | 1.03 |
| GII | 11.05 | XiS | 0.61 |
| X7500 | 6.57 | sigmaY | 0.61 |
| XiS | 6.02 | A | 0.61 |
| X1800 | 5.84 | E1800 | 0.61 |
| GiII | 5.09 | E7500 | 0.61 |
| epsilon | 3.11 | GiII | 0.61 |
| alpha12 | 2.47 | | |
| X1200 | 2.41 | | |
| SS | 2.36 | | |
| C | 2.07 | | |

Both FDR logworth and GSA (Sobol' indices) were performed with data generated from this RDSM (Table 11). Both are in agreement for the top three dominant parameters, *P*, *XS*, and *GiII*, with *GiII* replacing *GII* from the original analysis, exemplifying the importance of an adequate number of data points in the region under consideration. These top three influential parameters then comprised the final *DI* mechanism RDSM. Figure 14 shows the point cloud for these parameters from varying angles allowing for visualization of the significant interactions between the parameters. As the RDSM was formulated only for the engaged region, it was necessary to create a function based on the parameter ranges when *DI* is active. This plane was bound by a set of coordinates (*P*, *XS*, *GiII*): (0.4, 0, 0), (0, 0.5, 0), (0.85, 0.3, 1), and (0, 1, 1). If the data point fell on the left side of this plane, the RDSM reported a zero-energy value. If the data were on the right side, the *DI* RDSM is included in the total energy analysis.

**Table 11.** Focused FDR logworth values on the additional data representing the DI subspace.

| Variable | Total-Order Index | FDR Logworth |
| --- | --- | --- |
| P | 0.40 | 27.94 |
| GiII | 0.28 | 25.56 |
| XS | 0.22 | 18.68 |
| X1800 | 0.12 | 0.17 |
| C | 0.10 | 1.52 |
| X7500 | 0.08 | 1.52 |
| SS | 0.07 | 0.27 |
| XiS | 0.07 | 0.40 |
| X1200 | 0.05 | 0.66 |
| alpha12 | 0.05 | 0.81 |
| epsilon | 0.04 | 2.47 |
| GII | 0.04 | 1.16 |

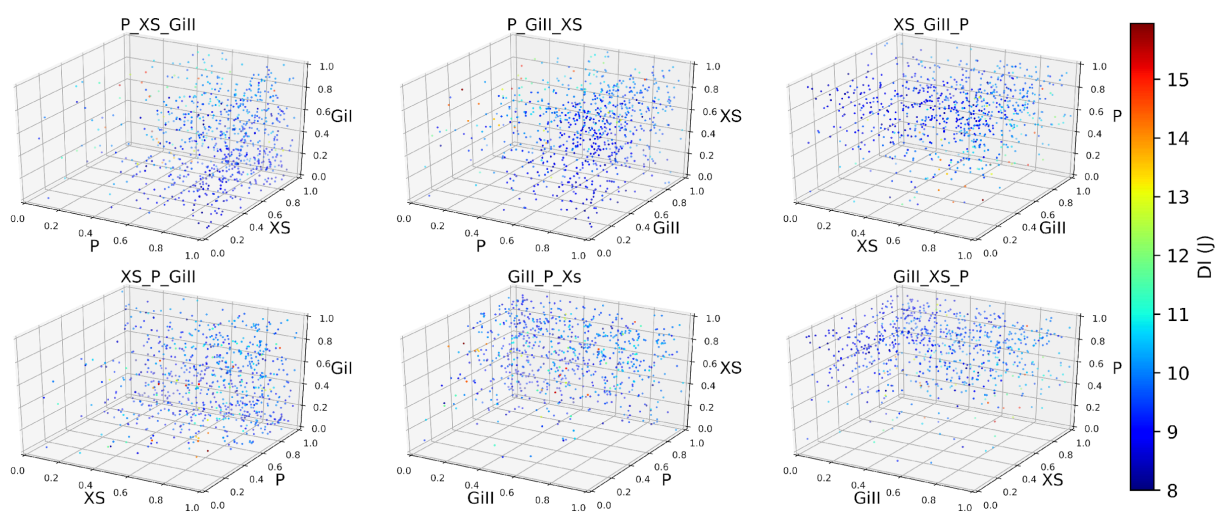

**Figure 14.** Scatter plots of the 3D *DI* subspace contoured with the energy absorbed due to disbond at the interface.

### 3.3. Comparison of the TS to the Summed Reduced-Dimension Surrogate Model

To evaluate the accuracy of the direct versus the summed approach, the final *TS* RDSM was compared to the summed RDSM (the summation of the mechanism RDSMs). (A linear summation inherently captured interactions between the damage mechanisms as all output energies were obtained simultaneously from the source model simulations.)

The *TS* RDSM was constructed with only 4 of the 41 input parameters (*A*, *E*, *XS*, and *Aln*), while the summed mechanism RDSM included 9 of the 41 input parameters (*A*, *E*, *Aln*, *X1800*, *E1800*, *X7781*, *P*, *XS*, and *GiII*). Both RDSMs were compared to the saved 25 data points from the initial FE predictions as well as an additional 200 data points generated separately for a total of 225 validation points (Table 12). The MAE of the summed approach was 5.78%, comparable to the 5.45% of the direct approach. For the validation points in the *DI* subspace, the summed RDSM outperformed the direct RDSM, with 4.63% and 5.51%, respectively. Figure 15a,b present the 45° validation plots using all 225 points. This result highlights that the summed approach is comparable in accuracy to the direct approach. Additionally, Figure 15c,d, showing the validation data in the engaged *DI* subspace only, clearly demonstrate the prediction improvement due to additional sampling in the mechanism subspaces during summed RDSM formulation.

**Table 12.** Comparison of direct and summed RDSM approaches.

| | **All Data Points** | | | *DI* **Important Data Points** | | |
|---|---|---|---|---|---|---|
| | **Total Energy Absorbed, *TS* (lbf-in)** | | | **Total Energy Absorbed, *TS* (lbf-in)** | | |
| | **FE** | **Source RDSM** | **Summed RDSM** | **FE** | **Source RDSM** | **Summed RDSM** |
| Mean | 247.37 | 241.48 | 238.54 | 249.54 | 241.26 | 245.62 |
| Std | 19.93 | 13.91 | 13.47 | 16.54 | 14.49 | 12.36 |
| %MAE | | 5.45% | 5.78% | | 5.51% | 4.36% |
| Std of MAE | | 3.91% | 3.98% | | 3.83% | 3.65% |

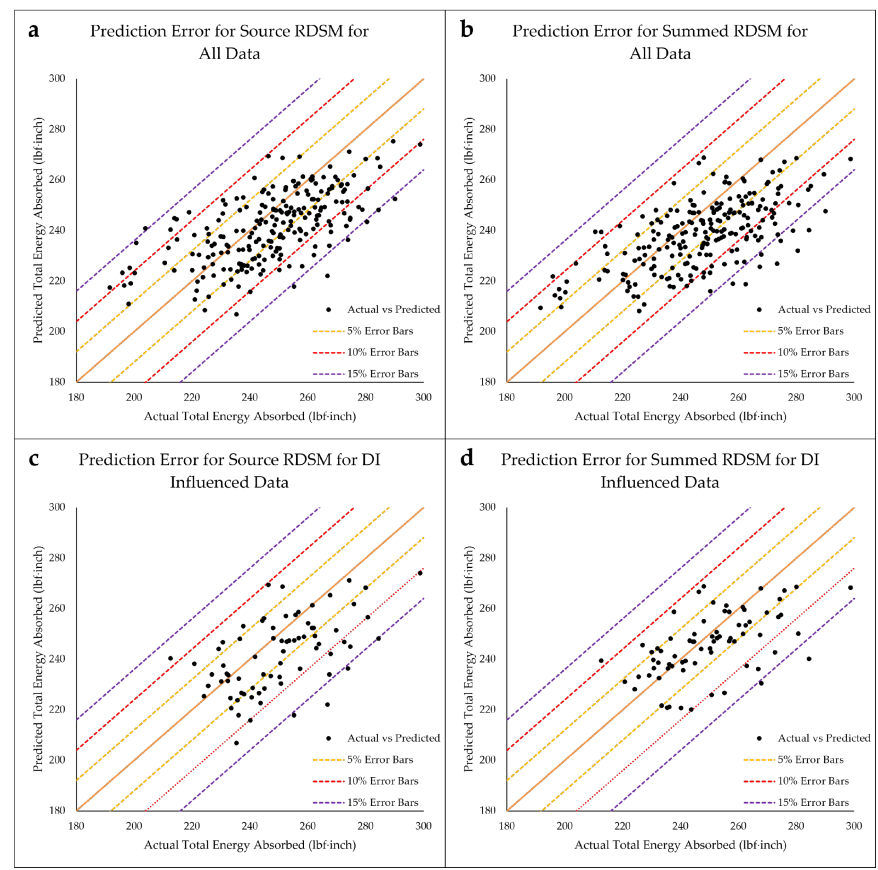

**Figure 15.** Error of 225 data points for the total energy absorption comparing the direct and summed RDSMs. (**a**) Direct approach in the entire parameter space, (**b**) summed approach in the entire parameter space, (**c**) direct approach for data points in the interface engaged subspace, (**d**) summed approach for data points in the interface engaged subspace.

## 4. Discussion

The results of this investigation illustrate that RDSM formulation in general provides an efficient and accurate method (within a specified error) to predict the output of high-fidelity models representing complex parameter spaces. Two approaches to RDSM formulation were investigated. The direct method formulates a RDSM of the total damage energy absorbed by the structure using the total energy extracted from the source model output. Alternatively, the summed approach formulates a mechanism RDSM for each damage type individually using the most dominant parameters for each mechanism. These mechanism RDSMs are then summed to obtain the RDSM predicting the total damage energy absorbed. The accuracy of both methods was comparable in general when considering the sampling of the entire parameter space of 41 variables. However, the summed approach exhibited significantly improved error reduction with the addition of data sampled directly within a mechanism subspace, as demonstrated for *DI*. While it is reasonably expected that the accuracy of both the direct and summed approaches will improve with additional data, the summed approach informs focused data collection in the parameter regions with the greatest influence on the output while simultaneously characterizing the damage mechanisms, offering a distinct benefit, in addition to a substantial reduction in computational time. This characterization of the individual damage mechanisms provides an understanding the complex, high-dimensional parameter space with more detail to inform optimized design, inspections, and experimental testing to improve the data quality of input parameters. Additionally, this understanding enables the design of composite/metal layered structures with tailored failure modes to induce a particular damage mechanism, for example, a failure mode that is readily detected as opposed to non-visible damage.

### 4.1. Direct Approach

The direct approach to RDSM formulation was computationally and manually expensive to generate, especially considering the need to develop a surrogate model that includes all 41 parameters as variables prior to parameter reduction with GSA. The subsequent GSA was also computationally prohibitive given the complex behavior of the source model and the large number of parameters. The use of a screening tool, such as FDR logworth, was demonstrated as a viable option for parameter reduction prior to surrogate model development. Comparison of GSA (Sobol' indices) and FDR logworth measures to evaluate parameter effects on the TS damage energy indicated that *A*, *E*, *XS*, *Aln*, *P*, and *GII* were the most influential parameters. Given that both methods predicted the same dominant parameters in the same order, FDR logworth was then successfully applied during the mechanism RDSM formulation for parameter reduction prior to surrogate modeling and GSA to evaluate parameter interactions.

The GSA for the *TS* damage energy indicated that for this specific configuration, the metal substrate was absorbing most of the damage energy in the system, thus reducing the impact of the other damage mechanisms on the *TS* damage energy as sampled across the entire parameter space. While this approach was shown to accurately predict *TS* damage energy throughout the overall parameter space, the summed approach provides more flexibility in accurately capturing behavior within subspaces where an individual damage mechanism is more significant, for example, the *DI* engaged subspace. The reduced impact of the other damage mechanisms in the direct approach was evident, as the GSA results for the individual damage mechanisms indicated many interactions between parameters deemed less influential than the metal substrate and matrix shear parameters. These other parameter combinations not included in the direct RDSM may significantly influence small subspaces which may not necessarily be captured or identified through general sampling of the entire parameter space.

### 4.2. Summed Approach

The summed approach enables individual characterization of the damage mechanisms providing insight into localized behavior. The shear plasticity (*PL*) damage mechanism was controlled by the properties of the E-LT 1800 fabric laminae and the matrix. The results show that a higher amount of energy was absorbed by matrix cracking if the E-LT 1800 had a combination of low tensile strength (*X1800*), high elastic modulus (*E1800*), and a high matrix shear strength (*XS*). The lower damage regions (Figure 12a) indicate that a strong interior 0°/90° fabric reduced the overall deflection of the specimen and reduced the overall stress on the resin, thus reducing the amount of energy absorbed by shear plasticity. The higher region indicates that a weaker interior fabric will cause the matrix to carry more of the load. Coupled with a strong resin matrix, more energy absorption by the *PL* damage mechanism can occur. Interaction between all three parameters was noted, indicating that matrix cracking absorbs more energy if the main fabrics of the laminate are stiffer and the matrix can resist shear.

Intralaminar fracture (*DL*) in the laminate, corresponding to fiber fracture, required high values of *XS*, *E1800*, and the tensile strength of the outer Hexcel 7781 fabric (*X7781*). Increasing the shear strength of the matrix allows the cohesive component to transfer load to the fabric and other laminae. This transfer is enabled by the matrix holding its bond longer and successfully distributing the load between the plies. Increasing the tensile strength of the outer fabric layer ensures that the maximum stress is maintained on a 0°/90° fabric. Delamination between layers within the composite laminate (*DC*) was shown to be controlled by only two parameters, *E1800* and *X1800*, with evident interaction between these parameters. These results indicated that a stiffer interior fabric, in this case the E-LT 1800 layers, restrains deformation reducing the interlaminar strain. Less-stiff plies coupled with a high strength will therefore absorb more damage energy as shown by the GSA results. The evaluation of the plastic deformation in the metal substrate (*PM*) mirrored the results from the *TS* parameter space given its dominance in the parameter space. *A*, *E*, and *Aln* controlled the damage mechanism parameter subspace as they did with the overall damage tolerance. No interaction between these parameters was identified. These results indicated that a stiffer metal substrate, higher yield strength, and low strain hardening allowed for more energy absorption.

The GSA results for the interfacial disbond (*DI*) damage mechanism indicated that this damage mechanism was significant in a localized parameter subspace and that the collection of additional data sampled in the engaged region efficiently increased prediction accuracy. The three dominant parameters were the shear hardening power term *P*, *XS*, and the shear fracture energy (*GiII*) of the interface and all exhibited significant interactions. The results for this damage mechanism also reinforced that the FDR logworth is an effective screening tool.

It is noted that the individual error of each RDSM for the parameter subspaces is higher than that of the *TS* surrogate model, excluding the *PM* RDSM. The error of the RDSMs could be improved with larger datasets or potentially considering a larger number of input parameters from the FDR logworth screening in the initial RDSM formulation. A significant advantage of the summed approach is that characterization of the damage mechanisms is part of the RDSM formulation. This process enables the identification of and focused data collection in localized subspaces that significantly contribute to the total damage energy. These subspaces may not necessarily be captured in the general sampling of a high-dimensional parameter space.

### 4.3. Future Improvements

Several aspects of the summed approach can be improved with future work including the implementation of more advanced damage modeling, generation of probability density functions for the most influential parameters, and development of weighted interaction functions. The FE source model included commonly used damage models integrated into Abaqus CAE including the CZM and a built-in Abaqus VUMAT for CDM. Emerging damage modeling approaches, such as NASMAT, could be integrated for higher resolution of damage initiation and propagation and to provide more flexibility to adapt to future advancements [76,77]. For example, NASMAT allows the user to localize the model down to its constituents and apply micromechanics to model damage at the material level. For this application, the composite damage would then include additional mechanisms such as fiber pullout and through-thickness damage. Additional understanding of the design parameters on the material level could be leveraged for NASMAT using microscopic imaging. However, procuring and analyzing the data would be time-consuming. Machine learning could provide an efficient approach to acquire the relevant data from the images. Recent advances in machine learning in Materials Science and Engineering have provided material scientists with new tools to expedite the collection of microstructure properties. These advancements include deep learning, programming question generation, and automated control and calibration [78-87]. HPC also continues to expand along with computing offload [88]. As the technology behind these systems deepens, so does the ability to perform these analyses with more efficiency and detail. Additional amounts of data could be quickly acquired to increase training data for machine learning along with more refined FE models that more realistically reflect the experimental tests.

The data quality of many parameters was poor as mechanical property population consisted primarily of manufacturer's data and was lightly supported by experimental testing. Increasing the quality of the data will provide better representation and prediction of the composite/metal layered structure's failure and controlling parameters. A suite of experimental tests should be performed to characterize the parameters included in the RDSMs. As experimental testing can be focused on a limited number of parameters, the development of probability density functions for these parameters would enable more advanced characterization and UQ. Improved parameter characterization can be compounded with additional experimental testing of the damage mechanisms and their interactions to inform and validate localized high-fidelity models. Weighting and/or interaction functions describing global behavior could then be populated using mechanism RDSMs formulated from these high-resolution models of the individual damage mechanisms.

## 5. Conclusions

The summed approach was demonstrated as a user friendly and efficient method to formulate RDSMs to predict not only the total damage energy but also that of each individual damage mechanism. Inherent to this approach is the characterization of the complex parameter space describing the progressive damage in the composite/metal layered structure under investigation. It was demonstrated that the summation of the mechanism RDSMs was able to predict the overall response of the structure within 5.78% using a small dataset. Another key finding was that the use of FDR logworth as a screening tool prior to preliminary RDSM generation is a viable method for parameter space reduction, thus greatly reducing computational time and resources. Ultimately, the damage tolerance of a composite/metal layered structure was predicted with RDSMs formulated using commonly available tools and techniques. While this paper focused on a particular composite/metal layered structure and loading configuration, the demonstrated approach is generally applicable to complex problems which can be broken into multiple components.

**Author Contributions:** Conceptualization, C.A. and S.T.; Methodology, C.A., B.H., and S.T.; Software, C.A. and C.C.; Validation, B.H. and C.A.; Formal Analysis, C.A. and S.T.; Investigation, C.A.; Resources, C.A. and S.T.; Data Curation, C.A., C.C., and R.B.; Writing—Original Draft Preparation, C.A. and S.T.; Writing—Review and Editing, C.C. and R.B.; Visualization, C.A.; Supervision, S.T.; Project Administration, S.T.; Funding Acquisition, S.T. All authors have read and agreed to the published version of the manuscript.

**Funding:** This research was funded by the Office of Naval Research (ONR), grant number N00014-21-1-2041.

**Data Availability Statement:** The dataset of 1555 points from the finite element (source model) analysis can be downloaded from the Harvard Database: https://doi.org/10.7910/DVN/QMC9HJ (accessed on 22 October 2023).

**Conflicts of Interest:** The authors declare no conflicts of interest. The funders had no role in the design of the study; in the collection, analyses, or interpretation of data; in the writing of the manuscript; or in the decision to publish the results.